\documentclass[conference]{IEEEtran}
\usepackage{subscript}
\usepackage{subfiles}
\usepackage{subfig}
\usepackage{graphicx}
\usepackage{caption}
\usepackage{amsmath}
\usepackage{amssymb}
\usepackage{multirow}
\usepackage{paralist}
\usepackage{algorithm}
\usepackage{algpseudocode}
\usepackage{setspace} 
\usepackage{xcolor,colortbl}

\begin{document}

\title{Sparse Matrix to Matrix Multiplication: A Representation and Architecture for Acceleration}

\author{\IEEEauthorblockN{Pareesa Ameneh Golnari\textsuperscript{1}}
\IEEEauthorblockA{Google Corp., pgolnari@gmail.com}
\and
\IEEEauthorblockN{Sharad Malik}
\IEEEauthorblockA{Princeton University, sharad@princeton.edu}
}

\maketitle

\begin{abstract}
% !TEX root = body.tex
Accelerators for sparse matrix multiplication are important components in emerging systems. In this paper, we study the main challenges of accelerating Sparse Matrix Multiplication (SpMM). For the situations that data is not stored in the desired order (row/column order), we propose a compact high performance sparse format, which allows for random access to a dataset with low memory access overhead. 
We show that using this format results in a 14-49 times speedup for SpMM. Next, we propose a high performance systolic architecture for SpMM, which uses a mesh of comparators to locate the useful (non-zero) computation. This design maximizes data reuse by sharing the input data among a row/column of the mesh.  We also show that, with similar memory access assumptions, the proposed architecture results in a 9-30 times speedup in comparison with the state of the art.

\end{abstract}

\begin{IEEEkeywords}
Sparse, Systolic Matrix Multiplier, SpMM, CRS
\end{IEEEkeywords}

\vspace{-10pt}
% !TEX root = body.tex
\section{Introduction}

In emerging data-inference applications the data-set{\let\thefootnote\relax\footnote{\textsuperscript{1} The author was with Princeton University when the work was done.}}
is often large and sparse. %These applications can be accelerated by skipping the access and computation for the zero elements. 
Hardware accelerators can benefit from sparse formats for these datasets as those store only the non-zero elements, reducing the required storage and bandwidth.
A common kernel in this context is sparse matrix multiplication (SpMM). 
%The need for accelerating SpMM is growing since 
This is used in many applications such as graph analysis, including breadth first search \cite{bfs} and graph clustering \cite{clustering}, in addition to its scientific applications in algebraic multi-grid solving \cite{multi-grid} and quantum modeling \cite{quantum}. 
While there are many accelerators suggested in the literature for sparse matrix to vector (SpMV) multiplication \cite{fpgauniversal, fpgastream, eric}, there are only a few proposed accelerators for SpMM. Accelerating SpMM cannot always be simplified to SpMV acceleration. As the second operand of SpMM is a two dimensional matrix rather than a vector. This poses additional challenges:

{\it Accessing data in two different orders:} In SpMM, the first dataset (the multiplicand) is accessed in row order and the other (the multiplier) is accessed in column order. However, sparse data formats store the non-zeros in one order, say row-order. Accessing data in the other orders has a high cost in the number of memory accesses (MA) and it might not be practical to store the large datasets in both orders. For instance, suppose matrix $A$ is stored in row order and appears as the first operand in $A \times B$. In another matrix multiplication, the same matrix $A$ might appear as the second operand, where it needs to be stored in column order.

 {\it Complexity of the SpMM algorithm:} Designing a high performance SpMM algorithm is challenging. SpMM algorithms should skip operations on zero elements. However, 
 in unstructured sparse datasets, location of zeros is arbitrary. This makes it non-trivial to locate the non-zeros without incurring an overhead that overwhelms the benefit of skipping zeros. This challenge will be elaborated in Section \ref{spmmsec}.

The first challenge, which is a special case of accessing sparse data in non-trivial patterns, has not been studied in the literature thus far. 
There are various high performance sparse formats proposed in the literature \cite{bitmap2, csr5, reduced6, eric}. However, most of these are proposed for SpMV acceleration, and therefore, they do not address the non-trivial access challenge.
% As mentioned earlier, the application of SpMM and consecutively the need for accelerating this operation is growing. 
On the architecture front, to the best of our knowledge, there is only one work to design a systolic-like structure for SpMM, the FPIC design~\cite{fpic}. In a conventional matrix multiplier (MM), there is a mesh of multiplier and accumulator (MAC) nodes and input data is shared among a row or a column of those nodes. This maximizes the reusability of data. 
%The previous FPIC SpMM design~\cite{fpic} performs independent input reading for each MAC node. Using extra buffering for input ports of the mesh nodes, they manage to lower the required bandwidth. However, their buffering method limits the size of the SpMM unit and the lack of scalability for this design increases the overall latency when it targets large matrices.

The previous FPIC SpMM design~\cite{fpic} does not have data sharing along the rows and columns, and each MAC node reads all its arguments directly from the inputs. To reduce memory bandwidth, these inputs are buffered at the rows and columns. This buffering limits the size of the SpMM unit and the lack of scalability increases the overall latency when it targets large matrices.

%independent memory accesses for each MAC node. While this focuses on the computation, it ignores the need for data resuse, and thus increases the required bandwidth.

%%%%%%%
In addressing the above gaps, this paper makes the following contributions: (i) We propose the Indexed Compressed Row Storage (InCRS) format, a new row-based sparse format with reduced memory access for arbitrary access patterns. We show how using InCRS can speedup SpMM when the dataset is not stored in the desired order. (ii)
We propose a high performance and scalable systolic SpMM. %that has advantages similar to the conventional systolic MMs while also targeting data reuse and thus reduced bandwidth.% 
In the proposed architecture, input is shared among a row or a column of the nodes and allows for maximum data reuse.

We evaluate our design on a range of large and sparse datasets. Using InCRS format when the second dataset is not stored in row-order, can speedup SpMM $\approx$ 14-49 times. Further, our systolic SpMM can speedup matrix multiplication 9-30 times in comparison with the state of the art FPIC design. %, which is resulted from $\approx$68-339 times reduction in memory access time of the computation.

% !TEX root = body.tex
\section{Complexity of Non-Trivial Data Access in Sparse Formats\label{spmmsec}}
%Unlike SpMV, the second operand in SpMM is a two dimensional matrix. 
In matrix multiplication the first matrix is accessed in row order and the second in column order. Since the second matrix might be accessed in row order in another application, e.g., being the first operand of another matrix multiplication, we assume all matrices are stored in one order, which we assume is row order (assuming column order is fine too, and 
will just switch the direction of our solution). 
Sparse formats store the non-zeros in one order, say row order, and accessing data in other orders is complex. For instance, to read one column of data stored in a row-based format, many of the non-zeros of each row are accessed to locate the elements of that column. In the following, we briefly study popular unstructured sparse formats and their cost of random data access.

\vspace{-5pt}
\subsection{Popular Unstructured Sparse Formats}

%The selected formats are all unstructured formats, which means they can store sparse datasets without restrictions on the sparsity patterns. 

\begin{enumerate}

\item {\tt ELLPACK}: One matrix stores the non-zero values of row $i$ in its $i^{th}$ row and another matrix stores the column indices of the non-zeroes~\cite{ellpack}.

\item List of lists ({\tt LiL}): A vector stores pointers to the beginning of each row. Then the non-zeros of each row are stored in a separate linked list. 

\item \addtocounter{enumi}{1} Co-ordinate list ({\tt COO}) and 4) Single linear list ({\tt SLL}): The non-zero values and their row and column indices are stored sequentially.
% {\tt COO} stores this data in a 2-D matrix and {\tt SLL} stores them in a linked-list or three separate vectors.

\item Jagged diagonal ({\tt JAD}): %The rows of the matrix are sorted by the number of non-zeros within them, in descending order~\cite{jad}. %and this order of the rows is stored in a vector. 
JAD is not row or a column order. It first sorts the rows of the matrix in descending order based on their number of non-zeros. Then it stores the non-zero values and their column indices starting with the first non-zeros of all the rows, then the second non-zeros of the rows and so on. One vector ($jadPtr$) stores pointers to the non-zeros of the first row~\cite{jad}.

%Then, the non-zero values and their column indices are stored starting with the row with the highest number of non-zeros, then the row with the second highest and so on. One vector ($jadPtr$) stores pointers to the non-zeros of the first row.
\item Compressed row storage ({\tt CRS}):
 CRS stores %the data in three separate vectors \cite{yale-crs}. 
 the non-zero values and their column indices in two vectors. Another vector provides pointers to the first non-zero of each row. %In this format $rowPtr[i]$ always gives the total number of non-zeros in the first $i$ rows of the matrix. Specifically, $rowPtr[M]$ gives the total number of non-zero values in the matrix.
Compressed column storage ({\tt CCS}) is the transpose of {\tt CRS}, where the matrix is stored in column order.

\end{enumerate}

\subsection{Cost of Non-Trivial Data Access}
Table \ref{memaccess} compares the complexity of reading {\itshape one} arbitrary element in sparse formats, where $M$ and $N$ are the number of rows and columns. $D$ is the density, i.e., the ratio of the number of non-zeros to the dataset size. Note that in conventional dense format, a single memory access is enough to read an arbitrary element.
%DONE-SM: Please use $D$ instead of $Dns$ for density.
\begin{table}[]
\footnotesize
\caption{Complexity of locating one data in sparse formats.}
\vspace{-7pt}
\begin{center}
\begin{tabular}{|c|c|c|}
\hline
Formats & MA complexity&  Avg \# of MA\\
\hline
{\tt ELLPACK},      & Access all the NZa before the & \multirow{2}{*}{ $\frac{1}{2}.N.D$}\\
{\tt LiL},{\tt CRS} &desired element in each row.& \\

\hline
 \multirow{4}{*}{{\tt JAD}}	&Search through the NZs of a row, one by& \multirow{4}{*}{ $N.D$}\\
				&one. Unlike in {\tt CRS}, the NZs of a row&\\
				&are not stored sequentially. Thus, locate &\\
				&each one of them using {\scriptsize $jadPtr$}.&\\
				
\hline
{\tt COO}, & There is no pointer. Access {\itshape all} the&\multirow{2}{*}{ $\frac{1}{2}.M.N.D$}\\
{\tt SLL}	& NZs located before the desired element. &\\
\hline
\end{tabular}
\end{center}
\label{memaccess}
\vspace{-9pt}
\end{table}
 As Table \ref{memaccess} shows, {\tt CRS} has amongst the least memory access requirements while it is also amongst the most compact formats.
Further, previous work \cite{date} has shown that {\tt CRS} incurs the lowest number of memory accesses in many linear algebra operations. However, {\tt CRS} still incurs a high number of memory accesses in comparison with the conventional dense format. 
Next, we propose a low-overhead improvement for accessing arbitrary data in {\tt CRS}.

% !TEX root = body.tex

\section{Accelerating Non-Trivial Data Access in Sparse Formats}

%As mentioned before, {\tt CRS }stores the non-zero values and their column indices in row order. It also stores pointers to the beginning of each row, which provides easy access to the rows. 
Suppose matrix $B$ is the second operand in an SpMM operation, then, SpMM needs to access elements of the columns of $B$ in order. To read element $B[i][j]$ in {\tt CRS} format, first the beginning of row $i$ is located, then all non-zero elements in that row ($B[i][:]$) and before column $j$ are accessed until $B[i][j]$ is located. This incurs an average of $\approx \frac{1}{2}.N.D$ memory accesses, which could be large for large datasets. For instance, for the Docword dataset from the NIPS 2013 challenge~\cite{ucidataset} with 12k columns and 4\% density, on average 240 memory accesses are required to locate  one arbitrary element. This makes reading a matrix in column order too slow.

We propose to augment {\tt CRS} by storing some information about the non-zeros in each row. 
Suppose we want to locate element $B[i][j]$. If we knew how many non-zeros exist before $B[i][j]$ in row $i$, say $n_{i,j}$ elements, then we could start from the beginning of the row $i$ and skip $n_{i,j}$ elements to locate $B[i][j]$ at the $(n_{i,j} + 1)^{th}$ location in the row. We propose the Indexed CRS ({\tt InCRS}) format, which augments the {\tt CRS} format with such information.

\subsection{Indexed CRS ({\tt InCRS}) Format}
 {\tt InCRS} divides each row into sections of $S$ elements, which are sub-divided into blocks of $b$ elements. It stores information on the number of non-zeros inside the sections and blocks. To access $B[i][j]$, this information is used to find the number of non-zeros that exist before the respective block to locate that block. This information is represented using counter-vectors, which are the addition to {\tt CRS}. {\it Each counter-vector is designed to be single word and stores information of one section. Therefore, to access the counter-vector of a section, only one memory access is required.} 

\begin{figure}[t]
\begin{center}
\vspace{-10pt}
\includegraphics[width=2in]{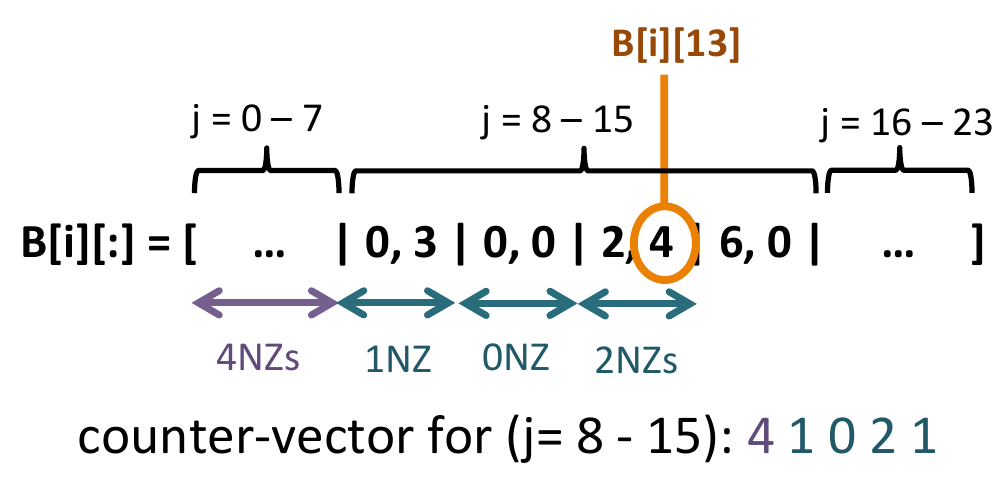}
\vspace{-10pt}
\caption{Example of a counter-vector for a section of a row.}
\vspace{-17pt}
\label{icrs_1}
\end{center}
\end{figure}

Figure \ref{icrs_1} shows an example of a row with 24 columns. In this setup, the row is divided to sections of 8 elements ($S = 8$) and each section is divided to blocks of 2 elements ($b= 2$). The first part of the counter-vector stores the total number of non-zeros that exist in the previous sections of that row and the rest of it stores information about the number of non-zeros {\itshape inside} each block in that section. In this example, the counter-vector indicates that there are between 5 to 7 number of non-zeros(NZs) located in row $i$ and before column $j=13$.

%DONE-SM: Explain the example in the figure in the text.

To locate $B[i][j]$, we first locate its counter-vector. $B[i][j]$ belongs to the section number $\lfloor{\frac{j}{S} \rfloor}$ and block number $\lfloor \frac{j \% S}{b} \rfloor$, where $\lfloor \frac{x}{y} \rfloor$ represents integer quotient of dividing $x$ by $y$, and $x \% y$ is the remainder of this division. Using the counter-vector, we find the number of non-zeros located before the $B[i][j]$'s block. %that contains column $j$. 
For this, we add the number of non-zeros that exist before $B[i][j]$'s section (first part of the counter-vector) to the number of the non-zeros that exist in the  blocks inside this section but before $B[i][j]$'s block. We then use this offset from the beginning of row $i$ to locate $B[i][j]$'s block and search through the elements of that block to locate $B[i][j]$. This means we can limit our search for the element $B[i][j]$ to only its block, which requires an average of $\approx \frac{b}{2} $ memory accesses. Recall that the counter-vector of each section is a single word and requires only one memory access. Therefore the average of overall memory accesses to locate a random element in this format is $\approx \frac{b}{2} +1$.

\subsection{Implementation of {\tt InCRS}}
%\subsubsection{Size of the sections and small blocks}
%As described above, {\tt InCRS} stores the distribution of non-zeros in each row to expedite locating an element in an array. 
In our implementation, the section size is 256 elements ($S=256$) and the blocks are 32 elements ($b=32$). 
%$S$ and $b$ being powers of two, simplifies finding the section and small block indices. 
In this implementation each counter-vector has 64 bits, where 8 $\times$ 6 bits are used for storing the non-zeros inside the 8 blocks. 
%To expedite extracting distribution information from this part of the counter-vector, especially for the small blocks closer to the end of the section, we incorporated an optimization method. However, this is not the focus of this paper since it does not affect the memory access requirements.
The remaining 16 bits of the counter-vector store the number of non-zeros before this section. This is based on the assumption that each row has maximum of 65k non-zeros, which is reasonable for a sparse dataset. Note that these parameters can be adjusted for a given dataset.

\subsection{Accelerating SpMM Data Access}
\label{benefit}
As mentioned before, the {\tt CRS} format requires on average $\approx \frac{1}{2}.N.D$ accesses to locate $B[i][j]$ and {\tt InCRS} reduces this to around $\frac{b}{2}+1$\footnote{Note that when the indices are sorted, we could benefit from applying binary search instead of linear search. However, CRS may not benefit in practice from binary search due to poor caching behavior as the accesses will not have locality. Thus, we applied the simpler linear search for both formats.}, which means the memory access reduces by a factor of $\frac{N.D}{b+2}$.
%\todo{Do you think this footnote is sufficient? or maybe even required?}
%WORKING- SM: Why will CRS not benefit from binary search for large datasets? In fact it should benefit even more. The reviewers may have a problem with this, especially as you are not also including the number of accesses for the Count-Vector.
%PG: I have initiated a new simulation for that. I will let you know as soon as I get any results on that.
This is the estimated reduction factor of memory accesses for reading one column of the dataset. Table \ref{benefitcost} shows this estimated factor for the 5 large and sparse datasets that we used in our experiments. The datasets are further discussed in Section~\ref{experiments}. The memory access ratio indicates that the datasets with a larger number of non-zeros in each row benefit more from using {\tt InCRS}. %For instance, Mks with average of 150 non-zeros in each row expects 3 times reduction in memory accesses when using {\tt InCRS}, while this reduction is 42 times for the Amazon dataset with the average of 1400 non-zeros in each row.  
%SM: What would this table look like if you used binary search and also included counting the number of accesses for the Count-Vector?

\begin{table}[]
\footnotesize
\caption{Cost and benefit of {\tt InCRS} compared to {\tt CRS}.}
\vspace{-5pt}
\hspace{-10pt}
\begin{tabular}{|c|c|c|c|c|c|}
\hline
\multirow{2}{*}{Dataset} & Dimension	& \multirow{2}{*}{$D$} & NZs per row& MA		& Storage\\
		& ($M \times N$)&		& (min, avg, max) 			&ratio			&ratio \\
\hline 
Amazon 	&300 $\times$10k &14\%	&(501, 1400, 2011)		&42			&0.99\\
\hline
Belcastro	& 370 $\times$ 22k& 6\%	&(1, 1300, 6787)		&39			&0.97\\
\hline
Docword	& 700 $\times$ 12k&4\%	&(2, 480, 906)			&14			&0.95\\
\hline
Norris	 	&1200 $\times$ 3.6k &1\%	&(3, 360, 795)			&11			&0.98\\
\hline
Mks		& 3.5k $\times$ 7.5k &1.5\%&(18, 150, 957)		&3			&0.88\\
\hline
%Snap		& 19k $\times$ 19k &0.0006&(0, 11,291)			&&\\
%\hline
\end{tabular}
\label{benefitcost}
\vspace{-5pt}
\end{table}%

The main cost of this improvement is the additional storage required for the counter-vectors. In our design, we require a 64bit counter-vector for each section (256 elements) of a row. Thus, this extra storage is $\frac{1}{S}.N.M$ words. Since the storage required for {\tt CRS} is $\approx 2.M.N.D$ words, the ratio of the storage required for {\tt CRS} format to the storage for {\tt InCRS} format is $\approx \frac{2D.S}{2D.S + 1}$, which we refer to as storage ratio in Table \ref{benefitcost}. As the Table shows, for these datasets, the storage required for {\tt CRS} varies between 0.99 and 0.88 of the storage required to store the same datasets in {\tt InCRS} format. It is clear that by reducing the size of the blocks the storage overhead (small $b$ is needed to fit the counter-vector in a word) and the expected benefit both increase. 

%The addition of the counter-vectors is a simple method to accelerate accessing one arbitrary address or non-trivial vector, say a column or a diagonal, of a dataset.
%A variation of this method can be also applied to some of the other sparse formats that provide easy access to the beginning of the rows, such as {\tt ELLPACK}. Moreover, it is easy to apply this method to the applications that already use formats such as {\tt CRS} or {\tt ELLPACK} since this method adds a separate storage to the initial format and does not change the way data is stored in those formats.
%SM: Deleted the above para for space.

% !TEX root = body.tex

\section{Systolic SpMM Architecture}
 
 In the previous section we proposed using the InCRS format to accelerate accessing the matrix elements for SpMM. In this section, we assume that data can be accessed fast enough and we focus on the SpMM architecture. 
 %In SpMM, since the zeros are skipped, the number of multiplications are considerably less than in conventional MM. However, the irregularity in the multiplication process complicates SpMM. 

\subsection{Main Challenges of Implementing SpMM}
\begin{figure}[t]
\vspace{-12pt}
\center
\hspace{-25pt}
\vspace{0pt}
\subfloat[Conventional systolic MM.]{\includegraphics[width=1.7in, height=1.3in]{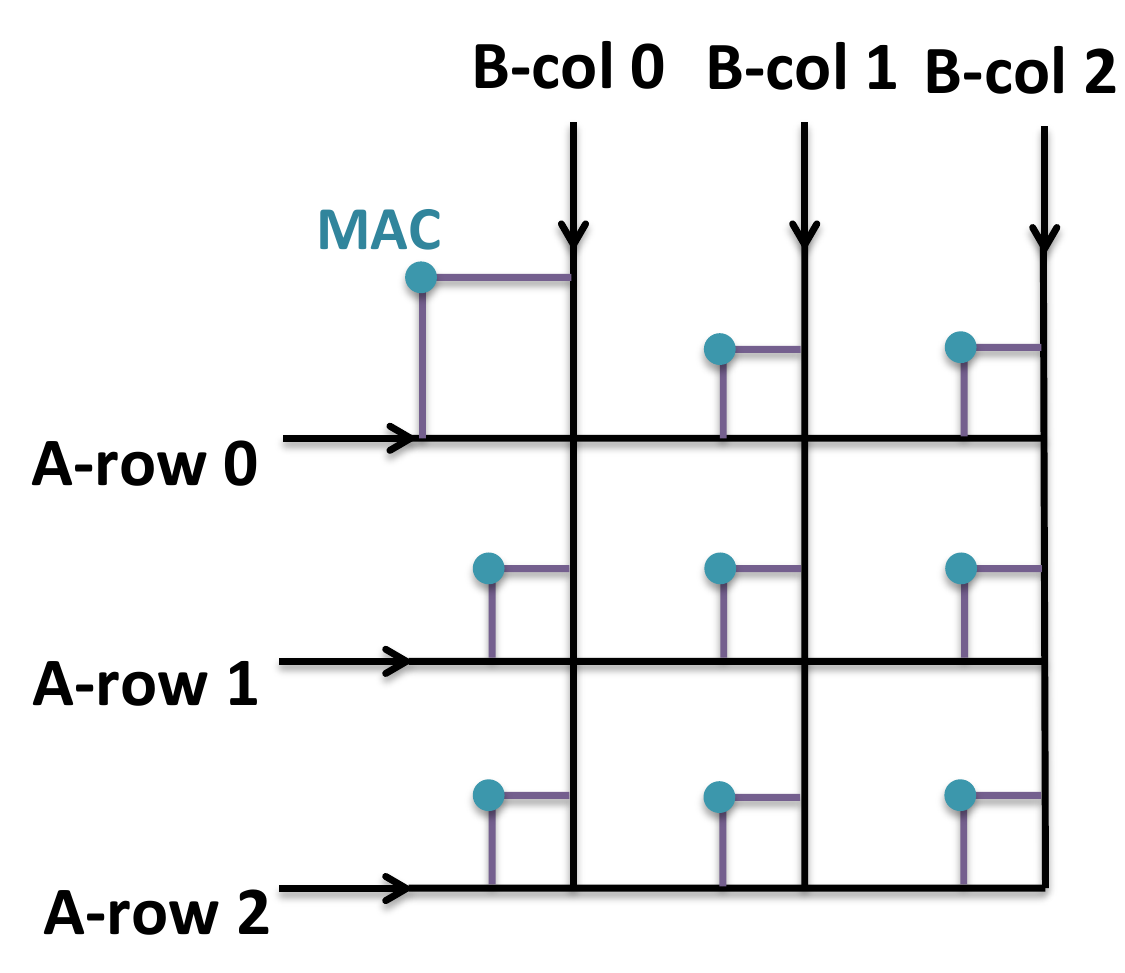}\label{systolic} }\hspace{8pt}
\subfloat[The proposed systolic SpMM.]{\includegraphics[width=1.7in, height=1.3in]{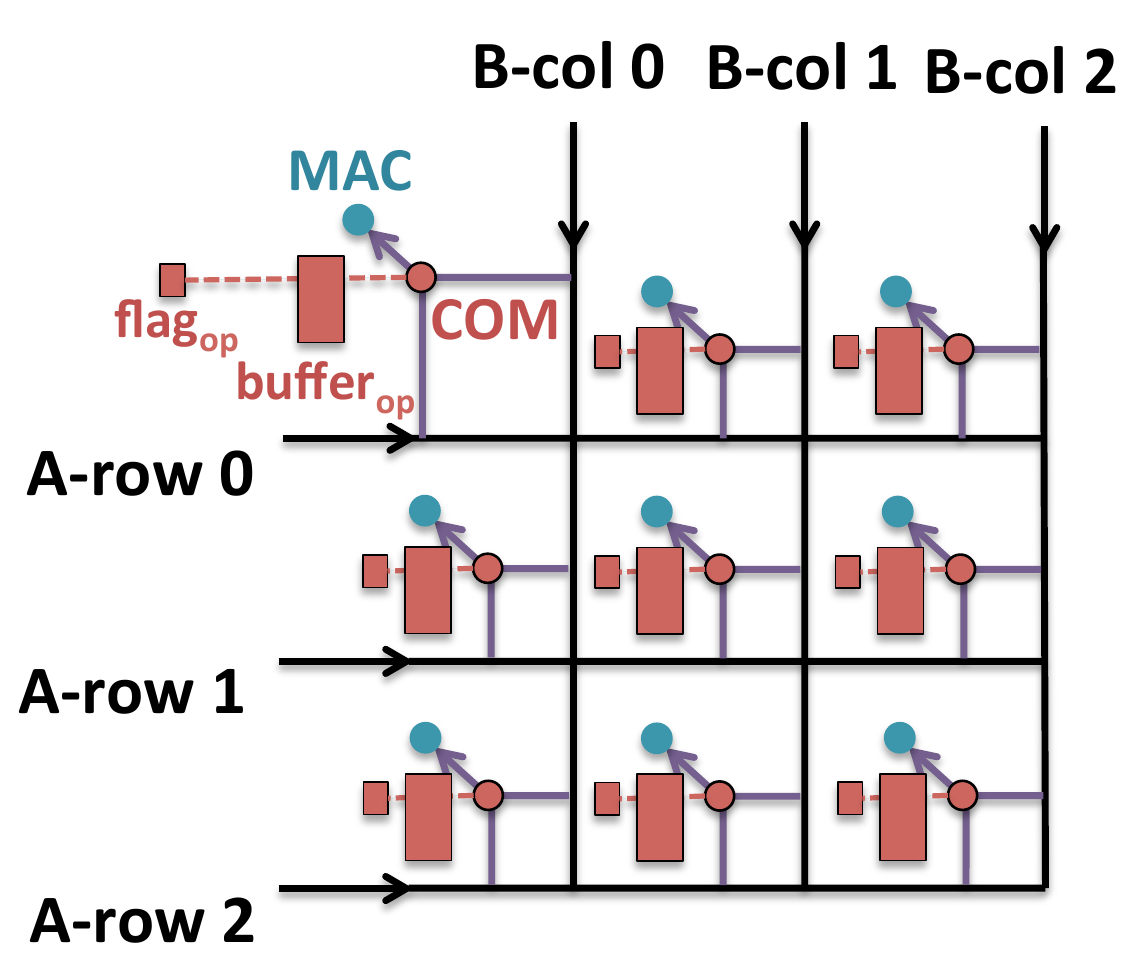}\label{mesh} }\hspace{-20pt}
\caption{$3\times 3$ systolic matrix multiplier.}
\vspace{-11pt}
\end{figure}

Figure \ref{systolic} depicts a $3\times3$ conventional systolic matrix multiplier, which we refer to as conventional MM. 

Each node in this mesh consumes two operands per cycle, either zero or non-zero, and performs a multiply-accumulate (MAC) on the pair. This gets more complicated in an SpMM architecture, since the nodes receive the data in sparse format. Each node in the SpMM case receives two sparse vectors (row of $A$ and column of $B$ in $A\times B$) and compares the indices, before MAC operations. Algorithm \ref{spmmalg} describes one cycle of the sparse dot product algorithm that each node of the mesh performs. %on the input sparse vectors.
Here, $c$ is the final product of the node. Vectors $a_{index}$ and $a_{val}$ hold the indices and values of nonzero elements in vector $a$, a row of matrix $A$. Similarly, vectors $b_{index}$ and $b_{val}$ hold the indices and values for vector $b$, a column of matrix $B$.

\begin{algorithm}
\small
\caption{One cycle of index-matching and MAC operation at SpMM mesh nodes.}
\begin{algorithmic}

\If {($a_{index}[i]$ = $b_{index}[j]$)}

\State $c \gets c + a_{val}[i] \times b_{val}[j]$
\State $i \gets i+1$
\State $j \gets j+1$
\ElsIf {($a_{index}[i] > b_{index}[j]$)}
\State $j\gets j+1$
\Else 
\State $i \gets i+1$
\EndIf

\end{algorithmic}
\label{spmmalg}
\end{algorithm}

%As psudocode \ref{convcomparison} shows, the comparator at each node compare the indices of the received vectors to find the matched indixes. 
In this comparison, if the indices are equal, MAC is performed on the operands and both counters $i$ and $j$ are incremented. When the counter of an operand is incremented, that operand is consumed and a new data can be read from the input. When the indices do not match, only the operand with smaller index is consumed while 
the operand with larger index is kept, i.e., either counter $i$ or $j$ is incremented. 
This asymmetric operand consumption results in a slower movement of operands than in conventional MM. Moreover, it results in stalling when the inputs are shared, i.e. the input is passed along multiple nodes along a row/column of the mesh. 
%For instance, one node might need to keep the received operand while the other node might need to drop the operand and read the next nonzero. 

%This arises two challenges. The first issue is that, each node in the mesh would consume 1 or 2 operands while in conventional MM each node consumes 2 operands at each cycle. The main two challenges in this comparison process is sharing the data among the comparators and consuming rate of each comparator.

%We overcome these challenges, i.e., low operand consumption rate and sharing operands among multiple nodes,  by synchronized move of the input data through the comparators.	
%%%%%%%
%\subsubsection{Independent Indices Comparison}
FPIC is the state of the art $8 \times 8$ systolic-like architecture proposed in \cite{fpic}. In that work, because of the discussed complexity in sharing the operands along a row/column, each node of the systolic architecture reads and compares the operands independently. This results in a high bandwidth requirement to memory, which is avoided in the FPIC design by buffering the input data. In this design, each node of the mesh reads from 8 possible buffers along the row, and from 8 possible buffers along the column, i.e. $8\times8=64$ buffers are used for reading data from matrix $A$ and 64 buffers to read data from matrix $B$. Each of these buffers is 32-element wide.
%each node of the mesh reads the input data from two buffers of 32-element wide, i.e., $8\times8=64$ buffers are used for reading data from matrix $A$ and 64 buffers to read data from matrix $B$. 
%\todo{I still don't understand the "two buffers of 32 elements wide." How about, "In this design, each node of the mesh reads from 8 possible buffers along the row, and from 8 possible buffers along the column, i.e. $8\times8=64$ buffers are used for reading data from matrix $A$ and 64 buffers to read data from matrix $B$." See if this is OK.}
%\todo{Sorry, I had made a mistake in describing the buffer's locations. Is it clear now?}

The main disadvantage of this design is that this buffering method limits the mesh size. In this work, they fixed the mesh size to $8\times 8$ and to scale the design, they suggest to use multiple $8\times 8$ units. 
%Moreover, FPIC does not solve the low operand consumption rate of Algorithm \ref{spmmalg}.
%\todo{But it does not need to because of its design.}
%\todo{PG: It is true that it does not need the symmetric operand consumption because of the Independence of the nodes. But still high operand consumption rate could have expedited the design.}
%the comparators perform the conventional sparse dot product algorithm described in Algorithm \ref{spmmalg}, which results in a lower operand consumption than conventional MM. 
In the following, we propose a synchronized mesh of comparators that overcomes the above-mentioned challenges.

%\subsubsection{Synchronized Move of the Data}
\subsection{Proposed Synchronized Mesh}
%We aim to design a systolic SpMM architecture, which skips zeros of the dataset while maintaining the nice features of conventional systolic MM. 
As mentioned above, the main advantages of the conventional systolic MM are: i) fast operand consumption, ii) sharing operands among a row or a column of nodes maximizing data reuse. 
We achieve this by a synchronized movement of the input data through the mesh of the comparators (Figure \ref{mesh}).
Each node of this mesh has a comparator, a buffer of operands and a flag, which are located before the MAC unit. 

\paragraph{Comparison process of the nodes}

\begin{algorithm}
\small
\caption{One cycle of index-matching and MAC operation at each node of the proposed synchronized mesh.}
\begin{algorithmic}[1]
\If {($a_{index}[i]$ = $b_{index}[j]$)}
\State $c \gets c + a_{val}[i] \times b_{val}[j]$
\State $reset (buffer_{op}, flag_{op})$

\ElsIf {($a_{index}[i] > b_{index}[j]$)}
\If {($flag_{op}$= $A$)}
\State  $(matched, val) \gets search (b_{index}[j], buffer_{op})$
\If {($matched$)}
\State $c \gets c + val \times b_{val}[j]$
\EndIf
\Else 
\State reset $buffer_{op}$
\State $flag_{op} \gets A $
\EndIf
\State $buffer_{op} \gets [a_{index[i]}, a_{val[i]}] $

\Else
\If {($flag_{op}$= $B$)}
\State  $(matched, val) \gets search (a_{index}[i], buffer_{op})$
\If {($matched$)}
\State $c \gets c + val \times a_{val}[i]$
\EndIf
\Else 
\State reset $buffer_{op}$
\State $flag_{op} \gets B $
\EndIf
\State $buffer_{op} \gets [b_{index[j]}, b_{val[j]}] $

\EndIf
\State $i \gets i+1$
\State $j \gets j+1$

\end{algorithmic}
\label{meshalg}
\end{algorithm}

As algorithm \ref{meshalg} describes, at each node, the comparator reads in two new operands, one along the row and one along the column, and compares their indices. If they match, the MAC operation is performed on them. If they do not match, it buffers the operand with the larger index (lines 14 and 25) and sets the $flag_{op}$ to indicate which operand ($A$ or $B$) has been buffered. %As the code describes, either $a$ operand or $b$ operand (or none) is stored at each node, which is indicated by $flag_{op}$. Therefore, one operand buffer is enough at each node. % as only $a$ or $b$ will need to be stored. 
The use of operand buffers allows for keeping the data moving and prevent stalling. Also, it allows for always consuming two operands (lines 27 and 28) rather than one or two operands per cycle at each node. This is possible since in the cases that operands' indices do not match, the operand with larger index is buffered instead of stalling, thus its counter still increments. Note that the buffer is not always occupied. It is flushed in some cases, e.g., when the indices of the operands match.

When the indices of the new operands do not match, the $flag_{op}$ is checked. 
If the operand with the smaller index can be matched with an operand residing inside the buffer, which is determined by checking the $flag_{op}$ at lines 5 and 16, the $buffer_{op}$ is searched. Here, $search(x,buffer)$ means searching $x$ among the operand indices stored in the $buffer$. The Boolean variable $matched$, indicates if the match was successful, and $val$, the value of the matched operand. %when $matched$ is true. 
Since the operand indices are sorted, $search$ operation requires at most $log_2(Depth_{buffer})$ comparisons. Considering small size of the buffers (32 elements here), we can use content addressable memories (CAM) to further accelerate  $search$.

%%%
\paragraph{Synchronization}
To prevent stalling, reading the rows and columns is synchronized at the end of each round of $R$ elements. This means that, at the $k^{th}$ round, each row or column accesses operands with indices between $k \times R$ and $(k+1) \times R$, then they wait for the rest of the rows and columns to finish the round. On starting a new round all the operand buffers are reset since any remaining buffer operands are no longer needed. This way, the operand buffers need to be at most $R$-element in size ($Depth_{op} = R$) to prevent any stall. 
In this work, we set $R$ to 32 elements. However, there is a trade off as %larger $R$ requires to synchronize the rows and columns less often. Therefore, 
larger $R$ allows for faster movement of the data because of less synchronization, at the cost of larger buffer overhead and search of the buffer elements.
%The main benefits of this architecture is that this mesh shares the rows of A and columns of B among the nodes of a row or a column similar to conventional systolic MM behavior and maximizes the data reuse. It also has a simple data access, avoiding extra buffering so it is possible to scale the architecture. 

Overall, the main advantages of this architecture are maximizing the data reuse and fast consumption of the operands. It also has a simple data access, avoiding extra buffering at the input ports, which makes the design scalable. The main drawbacks are the extra buffering and the search process through the buffer elements at each node. In the following, we evaluate this design and compare it with FPIC and conventional MM.

%\subsection{Benefit and Cost}
%- The main benefits of this architecture is that this mesh shares the rows of A and columns of B among the nodes of a row or a column similar to conventional systolic MM behavior and maximizes the data reuse. It also has a simple data access, avoiding extra buffering so it is possible to scale the architecture.

%- here we compare it with theFPIC method. in comparison with the other work, we do it much simpler. 

%also simplicity of data access and scalability. => larger blocks => higher data reuse and faster. Other benefit is using 2 operands, one from each direction, per cycle, while FPIC consumes 1 or 2 operands per cycle depending on the data.

%- In the experiments section we show that overall our architecture behaves better.

%- The main cost is the buffers- for blocks larger than 16*16 and comparisons, which could lower the clock frequency. 

% !TEX root = body.tex
\section{Experiments\label{experiments}}
We evaluated acceleration of SpMM memory access when using the InCRS format. Then, we evaluated the proposed systolic SpMM architecture.

\subsection{Methodology}

In the first set of experiments, our goal is to evaluate memory access acceleration when using the InCRS format for SpMM operation. 
For this, we used the Gem5 \cite{gem5} tool, which allows for simulating the memory hierarchy. %Our setup has a single core CPU working at 1GHz frequency and two levels of data caches. 
Table \ref{setuptable} summarizes the simulation parameters, which contains the typical values of memory size and other parameters.

\begin{table}[]
\caption{Gem5 simulation parameters.}
\vspace{-10pt}
\begin{center}
\footnotesize
\begin{tabular}{|l|l|}
\hline
CPU & Single core @ 1GHz frequency\\
\hline
L1 Data Cache	& 32kB, 2-way associative, LRU, Block Size=64 \\
L1 Instruction Cache & 32kB, 2-way associative, LRU\\
Hit Latency	& 2\\
\hline
L2 Cache	& 1MB, 8-way associative, LRU, Block Size=64\\
%Block Size & 64\\
Hit Latency & 20\\
Prefetching & Stride prefetching with degree 4.\\
\hline
\end{tabular}
\end{center}
\label{setuptable}
\vspace{-10pt}
\end{table}%
%\todo{PG: the L2 and L1 miss latency are not fixed parameters. They change for each simulation. For instance, they could depend on the number of page faults that occur.}
%SM: The L1 latency, L2 latency and page fault latency are all separately stated - these should be available and part of the table.}
In the second set of the experiments, we assume memory access is fast enough and the required data is ready at each clock cycle. Here, we aim to evaluate the latency of the SpMM algorithms. For this, we used cycle accurate simulation to model the accelerator's behavior.

\subsection{SpMM Memory Access Acceleration using {\tt InCRS}}

As mentioned in Section \ref{benefit}, the gain from using the InCRS format depends on the number of non-zeros per row. Therefore, we have selected datasets with a different average number of non-zeros per row ranging from 150 to 1400 per row (Table \ref{benefitcost})\cite{florida, ucidataset}. 
%We chose the detests that are more probable to be used for SpMM. For instance, graph and network datasets, or bag of words and user ratings datasets that are used by inference applications.
We select datasets such as graph and network applications, as well as bag of words and user ratings that are used by inference applications.

% lists these datasets in decreasing oder of the average number of non-zeros per row. 

We resized our target datasets because of the slow simulation speed in the experiments. We simplified the first operand (in SpMM) to a vector since the focus of this work is the column-order memory access to the second operand. Also, simplifying this part of the computation is fair because row-order access to a matrix is the same for the {\tt CRS} and the {\tt InCRS} formats. 
To resize the second operand, we randomly removed a number of rows from the datasets. At the end, the resized datasets were larger than at-least twice the L2 size to avoid eliminating the possible L2 cache miss effects. Table \ref{benefitcost} reports the sizes of the second operand after they are resized. We did not change or remove any of the {\itshape columns} of the second operand, since the columns' lengths and distributions of non-zeros are important factors in this comparison. 

\begin{figure}[t]
\vspace{0pt}
\begin{center}
\includegraphics[width=3in]{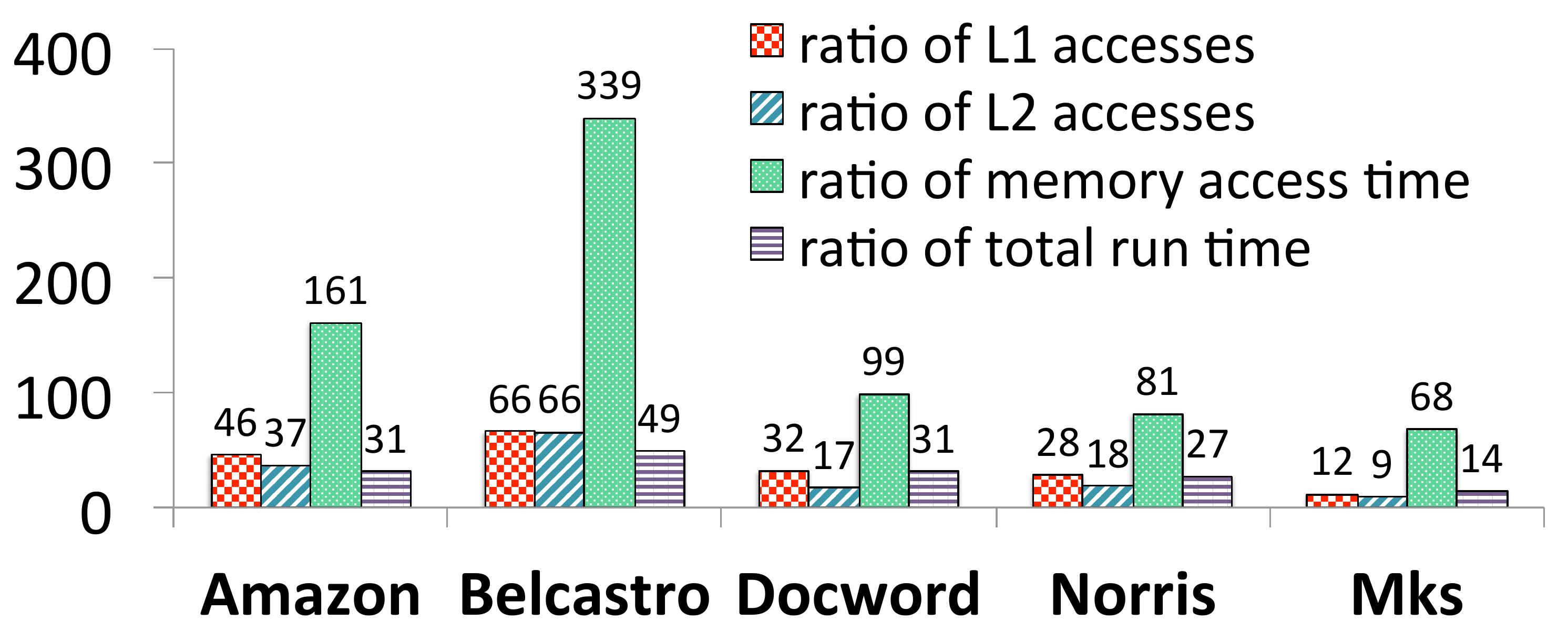}
\vspace{-5pt}
%\caption{The ratio of memory access parameters when using {\tt CRS} format to the same parameters when using {\tt InCRS}.}
\caption{{\tt CRS} vs. {\tt InCRS} memory access ratio}
\vspace{-20pt}
\label{swresults}
\end{center}
\end{figure}

Figure \ref{swresults} summarizes the benefits of replacing {\tt CRS} format by {\tt InCRS} for SpMM. 
This figure shows the number of cache accesses, total memory access time, and total run time when using {\tt CRS} format normalized to the same parameters when using {\tt InCRS}. As the figure shows, the number of cache accesses decrease for all the datasets with {\tt InCRS}. The ratios of memory access reduction for different datasets are close to our estimations in Table \ref{benefitcost}. For instance, the L1 access is decreased 49 and 31 times for Belcastro and Docword datasets, which is close to our estimation of 39 and 14 times reduction, respectively, in Table \ref{benefitcost}. Here, the average cache miss latency was 11k-22k cycles for L1 and 75k-91k cycles for L2.
%This is close to our expectation in Table \ref{benefitcost}, which expects more memory reduction for Docword than for Mks
%more memory reduction for Docword than for Mks
%we expected more memory access reduDocword dataset to ha  with ratios close to the expected benefits in Table \ref{benefitcost}. For instance, for Amazon dataset, the number of L1 accesses is decreased 46 times and as Table \ref{benefitcost} shows, we expected 42 time reduction in the number of memory accesses.

The third column in Figure \ref{swresults} shows the ratio of the overall time spent on memory accesses. %which is calculated based on the number of cache misses and hits and the respective latency. 
%This amount of reduction in the number of memory accesses results in reduction in the overall memory access time. The third 
As a result of memory access speedup, the overall execution time of SpMM decreases noticeably for all the datasets. For instance, SpMM on Docword is around 31 times faster if we use {\tt InCRS} instead of {\tt CRS}.

%%%\todo{may be I should add the number of hardware prefetching and L2miss rate to show that this format is also benefiting from prefetching}

\subsection{Evaluation of the Systolic SpMM Architecture}
We compare the performance of our proposed architecture with the conventional MM and the FPIC work \cite{fpic} performing $A\times A^T$ on a set of large and sparse datasets with a range of different densities (Table \ref{dataset2}). 
\begin{table}[]
\caption{Datasets used in the order of their densities.}
\vspace{-10pt}
\begin{center}
\footnotesize
\begin{tabular}{|c|c|c|c|c|c|c|c|}
\hline
Dataset&\textbf{Amazon}  & \textbf{Docword}& \textbf{Mks} & \textbf{Norris}\\ 
\hline
$D$&14\%&4\% &1.5\%&1\%\\
\hline
Dimension& $1.5k\times 10k$& $1.5k\times 12k$ &$7.5k \times 7.5k$ & $3.6k\times 3.6k$\\
\hline
\hline
Dataset& \textbf{Arenas} & \textbf{Bates} & \textbf{Gleich} & \textbf{Sch}\\
%\multicolumn{2}{c|}{Gleich}\\
\hline
$D$&0.85\%&0.11\%&0.095\%&0.057\%\\
%\multicolumn{2}{c|}{0.00095}\\
\hline
Dimension &$1k\times 1k$ & $2.5k\times 2.5k$&$2.6k\times 2.6k$ &$3.6k\times 3.9k$ \\
%\multicolumn{2}{c|}{$2.6k\times 2.6k$} \\
\hline

\end{tabular}
\end{center}
\vspace{-17pt}
\label{dataset2}
\end{table}%
The FPIC unit size is fixed to $8\times 8$ in \cite{fpic} %since this results in the fastest architecture 
and to scale this design, the authors suggest to use multiple units. Since they do not provide details about how to schedule the computation among the multiple units, we assume the best case scenario, which is perfect load balancing among the units. Based on that, to estimate the latency of the computation when using $k_{FPIC}$ numbers of the $8\times 8$ units, we performed the computation using a single unit and divided the latency of the computation by $k_{FPIC}$. 

Our design is different from FPIC in terms of the required bandwidth and buffers. To perform a fair comparison, we fix each of those two parameters one at a time and sweep the other one. As the focus of the following experiments is memory access and index-matching of SpMM, for simplicity we assume a single cycle latency for all operations including MAC and comparisons. This is the same for the conventional MM, FPIC, and the proposed SpMM.%and compare the overall latencies of the designs.

{\it Similar input bandwidth:}
The left %hand 
side of equation \ref{bweq} %indicates 
is the required bandwidth for our proposed design with a mesh of $N_{synch} \times N_{synch}$ nodes. The right side is the bandwidth required by FPIC, where $k_{FPIC}$ is the number of $ 8\times 8$ units, and $W_{tot}$ is the matrix elements' widths. Here, we assume the index width is 16 bits and value width is 32 bits.

\vspace{-9pt}
\begin{equation}
\label{bweq}
{2 \times N_{synch}} \times W_{tot} = 2 \times 8 \times k_{FPIC} \times W_{tot}
\end{equation}

{\it Similar buffer sizes:}
The left side of equation \ref{buffeq} indicates the number of 32-element buffers required for the proposed method and the right side indicates this number for FPIC. 

\vspace{-6pt}
\begin{equation}
\label{buffeq}
N_{synch}^2  = 2 \times 8^2 \times k_{FPIC}
\vspace{-6pt}
\end{equation}

%In this comparison we used one dataset with high density (Amazon) and one dataset with low density (Gleich). 
%Figure \ref{fixbw} and \ref{fixbuff} compare the overall latency of our proposed architecture with FPIC for the same bandwidth and buffer requirements respectively.

%%%figure was here

\begin{figure}[t]
\center
%%%\vspace{-10pt}
\hspace{-25pt}
\subfloat{\includegraphics[width=1.4in]{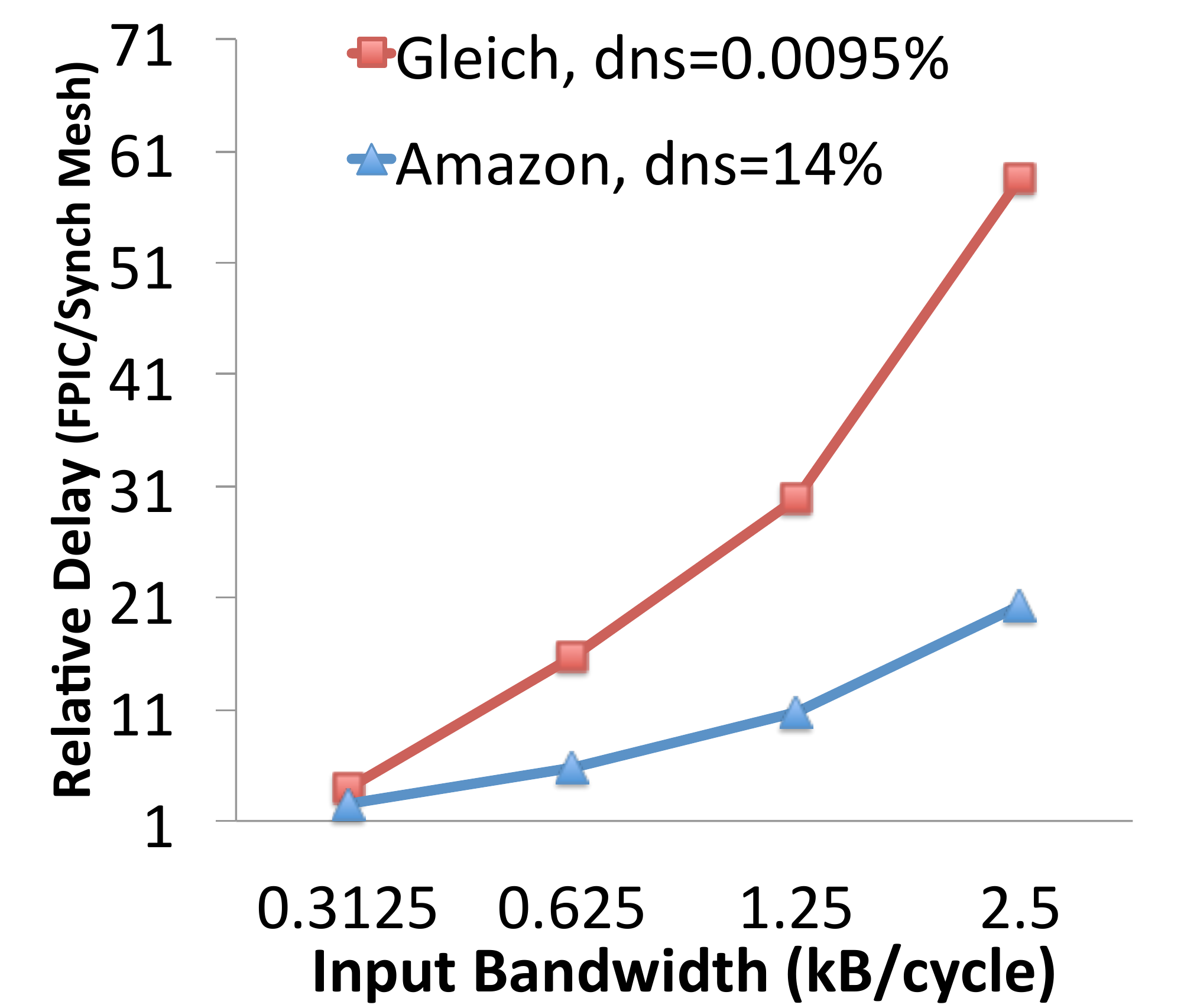}\label{fixbw} }\hspace{-6pt}\subfloat{\includegraphics[width=1.4in]{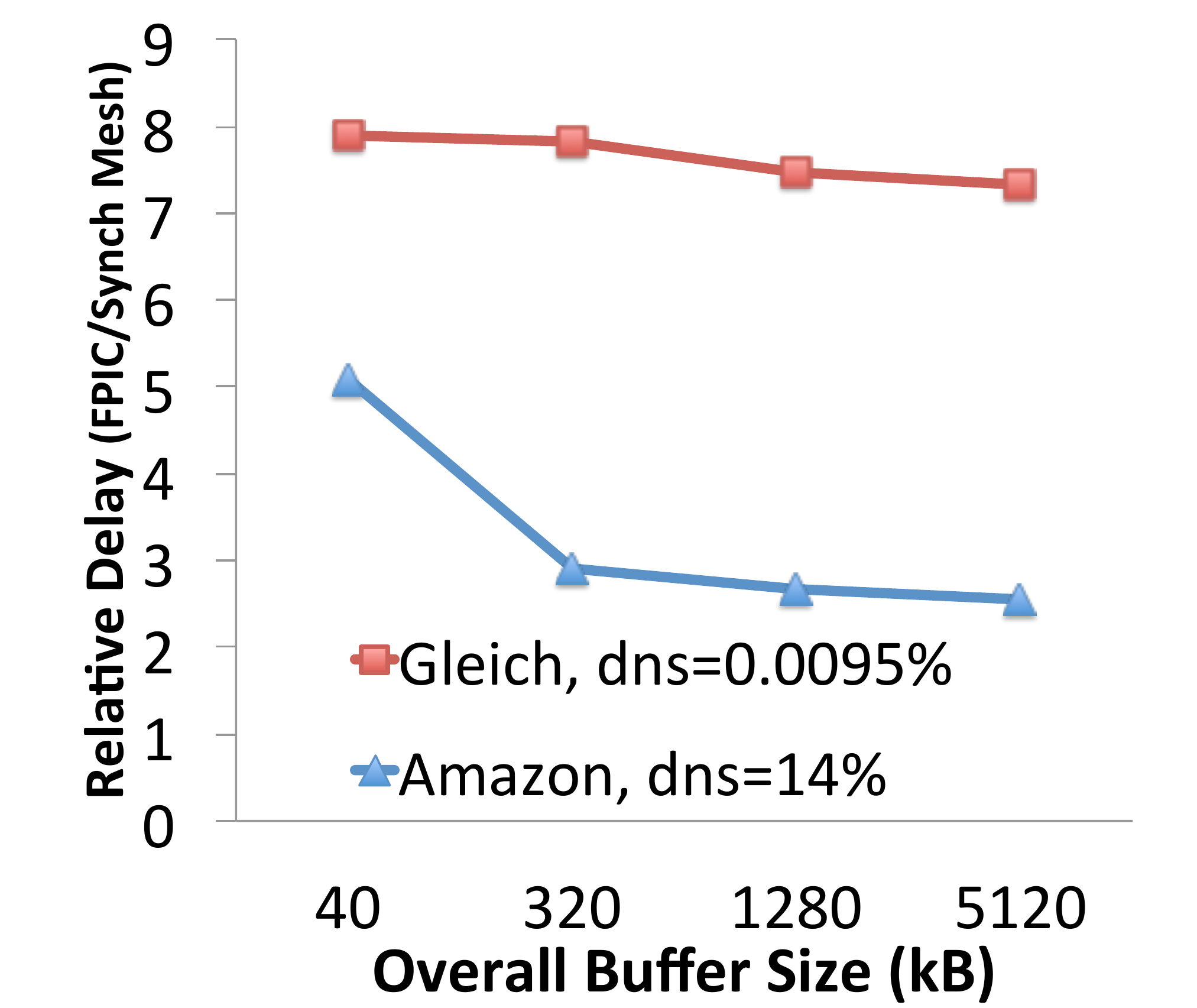}\label{fixbuff} }\hspace{-8pt}
%\caption{The ratio of the FPIC's overall latency to the synchronized mesh's overall latency.}
\caption{Comparison of the overall latency}
\vspace{-17pt}
\end{figure}

%As the figure shows, the main advantage of the proposed synchronized mesh to the FPIC method is the scalability of this design that allows for selecting a larger block in matrix multiplication and decreasing the bandwidth requirements. 
Figure \ref{fixbw} shows that assuming the same input bandwidth, the synchronized mesh performs SpMM 2.5-20 times faster than FPIC for the high density dataset and 4-58 times faster for the sparser dataset. The main disadvantage of this design in comparison with FPIC is its higher overall buffer size. However, Figure \ref{fixbuff} shows that even with the same overall buffer size and, consequently, lower bandwidth usage, the synchronized mesh still performs SpMM faster than FPIC for both the low and the high density datasets.

In the last set of experiments, we fixed the parameters for our design: $N_{synch}$ is set to 64, resulting in 4096 MAC nodes, which is ~60\% of floating point DSP slices available on the new {\small Xilinx FPGA (Virtex Ultra Scale XCVU9P} \cite{xilinx}). The rest of the parameters are set accordingly (Table \ref{parameters}). Here, we selected reasonable values for our parameters based on FPGA resources and we performed cycle-accurate simulation to evaluate the design. In the future, we will study the parameter selection process in more detail and implement the design on FPGA.

\begin{table}[t]
\caption{SpMM design parameters.}
\hspace{-7pt}
\vspace{-4pt}
%\begin{center}
\footnotesize
\begin{tabular}{|c|c|c|c|c|}
\hline
\multirow{2}{*}{Designs} & \multirow{2}{*}{\#Units, $N\times N$} & BW & \multirow{2}{*}{\#MACs} & Buffer\\
&&(kb/cycles)&&(kB)\\
\hline
\textbf{This work} & 1, $64\times 64$ &6 & 4096 &768\\
\hline
\textbf{FPIC-same BW} & 8, $8\times8 $&6 & \cellcolor{cyan} 512&192\\
\hline
\textbf{FPIC-same buffer} &32, $8\times8$& \cellcolor{red} 24& 2048 &768\\
\hline
\textbf{Conv. MM} &1, 96$\times $96&6 & \cellcolor{red} 9216 &-\\
\hline

%\multicolumn{2}{c|}{$2.6k\times 2.6k$} \\
\hline

\end{tabular}
%\end{center}
\vspace{-5pt}
\label{parameters}
\end{table}%

%The required input bandwidth is 768 B/cycle and overall buffer size is 768 kB. Assuming the same bandwidth requirement for the conventional MM, we set $N_{conv}$=96 since $N_{conv} = \frac{W_{tot}}{W_{val}}\times N_{synch}$. And for the FPIC design, we set $k_{FPIC}$= 8 (same BW) and $k_{FPIC}$=32 (same buffer size).

\begin{figure}[t]
\begin{center}
\hspace{-20pt}
\includegraphics[width=3.5in]{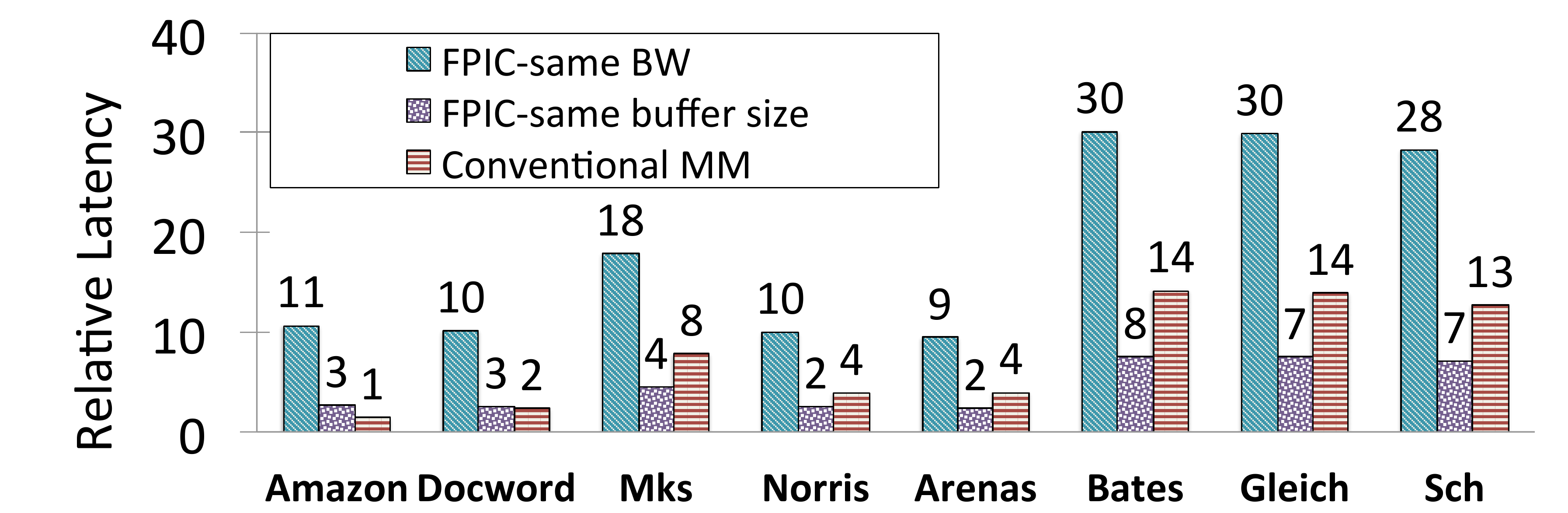}
\caption{Latency of the different architectures normalized to the latency of the proposed synchronized mesh.}
\label{spmmres}
\end{center}
\vspace{-20pt}
\end{figure}

Figure \ref{spmmres} compares the overall latency of the three designs for datasets with 0.057\% to 14\% density. Overall, our architecture's acceleration increases as the density decreases. For this range of densities, the proposed architecture performs SpMM 1.5-39 times faster in comparison with the conventional MM and 2-30 times faster in comparison with FPIC. The lower performance of ``FPIC-same-BW" design is a result of low DSP utilization and input BW (blue cell in Table \ref{parameters}).
%DONE-SM: Clarify what you mean by imbalance here.
%The colored cells in Table \ref{parameters} highlight the undesired design parameters. 

%The red cells in Table \ref{parameters} highlight the high resource requirement that might result in slowing down the design. 
The red cells in Table \ref{parameters} highlight the high resource requirement that might not be practical.
%In case this amount of resource is not available on the FPGA, the design would be slowed down.  
%On the other hand, the high amount of BW requirement for the ``FPIC-same-buffer" design and the high number of MACs required by $96\times96$ conventional systolic MM (red cells in the Table), can make those design choices impossible. More realistic designs would have smaller sizes, which have higher latencies. 
For instance, the overall good performance of the conventional MM in these experiments is a result of large $N_{conv}$, which is set based on the $N_{conv}=\frac{W_{tot}}{W_{val}}\times N_{synch}$ equation assuming the same input BW for this design and the proposed SpMM. However, implementing this size of mesh is possible only if there are enough DSP resources available on the host FPGA.

%Overall, the main advantage of the proposed synchronized mesh to the FPIC method is the scalability of this design that allows for selecting a larger block size in matrix multiplication and decreasing the bandwidth requirements. While FPIC design solves the bandwidth issue for $8\times8$ blocks, this design is not scalable and this limits its performance. 

% !TEX root = body.tex
\section{Conclusion}
Matrix multiplication is a key kernel in inference applications, which often have large and sparse data sets. Thus, there is strong interest in accelerating this important operation with sparse data representations. We studied the main challenges of accelerating Sparse Matrix Multiplication (SpMM) including non-trivial data access and designing a systolic architecture. We proposed a modification to sparse formats to accelerate access to a dataset when the dataset is not stored in the desired order (row/column order). Our experiments show that the proposed InCRS format speeds up SpMM 14-49 times due to fewer memory accesses. Next, we proposed a high performance systolic SpMM architecture, which accelerates the computation and maximizes the data reuse by performing synchronized movement of the data through the %rows and columns of the nodes of the systolic 
architecture. We showed that, with similar memory access assumptions, the proposed architecture performs SpMM 9-30 times faster than the state of the art FPIC architecture.

\section*{Acknowledgment}
This work was supported in part by {\small SONIC}, one of the six {\small SRC STARnet} centers.
\bibliographystyle{abbrv}
\footnotesize \bibliography{bibFile}

\end{document}